\documentstyle[aps,prl,epsf]{revtex}
\def\ba{\begin{eqnarray}}
\def\ea{\end{eqnarray}}

\newcommand{\eqb}{\begin{equation}}
\newcommand{\eqe}{\end{equation}}
\newcommand{\dmb}{\begin{displaymath}}
\newcommand{\dme}{\end{displaymath}}
\newcommand{\pd}{\partial}
\newcommand{\ep}{\varepsilon}
\newcommand{\eab}{\begin{eqnarray}}
\newcommand{\eae}{\end{eqnarray}}
\newcommand{\ra}{\right\rangle}
\newcommand{\la}{\left\langle}

\newcommand{\nc}{\newcommand}
\nc{\lab}{\label}
\nc{\eq}{Eq.\,(\ref}
\nc{\eqs}{Eqs.\,(\ref}
\nc{\tm}{\tiny\mbox}
\nc{\vivi}{very interesting and very important}
\nc{\al}{\alpha}
\nc{\ga}{\gamma}
\nc{\de}{\delta}
\nc{\ze}{\zeta}
\nc{\et}{\eta}

\nc{\Th}{\Theta}
\nc{\ka}{\kappa}
\nc{\lam}{\lambda}
\nc{\rh}{\rho}
\nc{\si}{\sigma}
\nc{\ta}{\tau}
\nc{\up}{\upsilon}
\nc{\ph}{\phi}
\nc{\ch}{\chi}
\nc{\ps}{\psi}
\nc{\om}{\omega}
\nc{\Ga}{\Gamma}
\nc{\De}{\Delta}
\nc{\La}{\Lambda}
\nc{\Si}{\Sigma}
\nc{\Up}{\Upsilon}
\nc{\Ph}{\Phi}
\nc{\Ps}{\Psi}
\nc{\Om}{\Omega}
\nc{\ptl}{\partial}
\nc{\del}{\nabla}
\nc{\be}{\begin{equation}}
\nc{\ee}{\end{equation}}
\nc{\bea}{\begin{eqnarray}}
\nc{\eea}{\end{eqnarray}}
\nc{\ov}{\overline}
\nc{\gsl}{\!\not}

\draft

\begin{document}

\title{Confinement, Chiral Symmetry Breaking, and Axial Anomaly 
from Domain Formation at Intermediate Resolution}

\author{Ralf Hofmann}
\address{Max-Planck-Institut f\"ur Physik\\ 
Werner-Heisenberg-Institut\\ 
F\"ohringer Ring 6, 80805 M\"unchen\\ 
Germany}

\maketitle

\centerline{ 
MPI-PhT  2001-07\\ 
May 2001 
}

\begin{abstract}

Based on general renormalization group arguments, Polyakov's 
loop-space formalism, and recent analytical 
lattice arguments, suggesting, after Abelian gauge fixing, a description 
of pure gluodynamics by means of a 
Georgi-Glashow like model, the corresponding 
vacuum fields are defined in a nonlocal way. 
Using lattice information on the 
gauge invariant field strength correlator in full QCD, 
the resolution scale $\La_b$, at which these fields 
become relevant in the vacuum,  
is determined. For SU(3) gauge theory it is found that 
$\La_b\sim$ 2.4 GeV, 3.1 GeV, and 4.2 GeV for ($N_F=4,\,m_q=18$ MeV), 
($N_F=4,\,m_q=36$ MeV), and pure gluodynamics, repectively.   
Implications for the operator product expansion 
of physical correlators are discussed. It is argued that 
the emergence of magnetic (anti)monopoles in the vacuum at 
resolution $\La_b$ is a direct 
consequence of the randomness in the 
formation of a low entropy Higgs condensate. This implies a breaking of 
chiral symmetry and a proliferation of the axial U(1) anomaly 
at this scale already. Justifying Abelian projection, a 
decoupling of non-Abelian gauge field fluctuations 
from the dynamics occurs. The condensation of (anti)monopoles at $\La_c<\La_b$ follows 
from the demand that vacuum fields ought to have vanishing action at 
any resolution. As monopoles condense they are reduced 
to their cores, and hence they become massless. Apparently broken gauge 
symmetries at resolutions $\La_c<\La\le\La_b$ are restored in this process.

\end{abstract} 



\section{Introduction}

The question of how confinement happens in Quantumchromodynamics (QCD) \cite{QCD} is an old one. 
An appealing proposal due to 't Hooft and 
Mandelstam \cite{HM} is 
the dual superconductor picture 
of the QCD vacuum. In this scenario the 
formation and subsequent condensation 
of Abelian magnetic monopole degrees of 
freedom at low resolution leads to a 
linearly confining potential 
between heavy color charges for 
distances larger than the resolution. 
Unfortunately, in QCD there are nor fundamental 
monopoles neither are there classical solutions of finite action exhibiting 
long-lived magnetic monopoles 
\cite{Brower}. A possibility 
of defining magnetic monopoles by 
the point-like singularities of a (partial) 
Abelian gauge fixing \cite{'thooft2} 
renders the monopoles to be gauge variant objects. 
On the other hand, lattice 
simulations working in Abelian gauges and subsequent projection onto 
Abelian fields reproduce the bulk of the tension $\sigma$ 
of the confining string between 
static color charges. Moreover, within Abelian projection it is claimed 
that the generation of the string tension 
is mostly due to the monopole dynamics. The purpose of 
this work is to develop a framework in which ad hoc Abelian 
projection is not needed to explain the 
low energy features of QCD. Rather, it will turn out to be 
a {\sl consequence} of the dynamics of suitable vacuum fields. 
Thereby, the old idea of a dynamical origin of Higgs fields 
in pure gauge theories \cite{Kleinert}, general 
renormalization group arguments, and the recently advocated expressibility 
of Abelian gauge fixed pure lattice QCD in 
terms of an adjoint Higgs model \cite{Sur} serve as motivations. To define a chiral field 
and an adjoint scalar (hencefore referred to as Higgs field), which are relevant for the 
vacuum dynamics at intermediate resolution, we appeal to 
Polyakov's loop-space approach to gluodynamics \cite{Pol2}. 
Both fields are defined in a nonlocal way as functional 
integrals over (degenerate) loops with a common base point $x$. The extension of these 
loops is determined by their invisibility at resolution $\La_b$, that is, the locality of the 
so-defined fields. The resulting field 
possesses zero curvature. In order to find the scale $\La_b$ at which classical Higgs-field 
configurations become relevant in the vacuum one can make contact with the gauge invariant 
field strength correlator \cite{Dosch}. This quantity has been measured in full 
QCD and SU(3) gluodynamics on a lattice \cite{DiGiacomo,Me}. The formation of magnetic monopoles, and hence 
the onset of chiral symmetry breaking \cite{Stern} and the axial U(1) anomaly, 
are argued to be a consequence of the randomness of Higgs condensation at 
this scale. Condensation of monopoles is observed with probes of momentum $\La_c<\La_b$. It 
follows from the ignorance of translational symmetry breaking in relevant, classical configurations 
at higher resolution and the fact 
that vacuum fields viewed at {\sl any} resolution have to have a small action. 
Monopoles then are reduced to their 
cores, and hence they become massless. As already stated by Polyakov \cite{Pol}, 
the process of monopole condensation (and hence the 
confinement of color charge) goes together with the restoration of the 
apparently broken gauge symmetry.     
 
The paper is set up as follows: 
In the next section nonlocal definitions of vacuum describing fields in terms of fundamental field strength 
are made. Thereby, the fundamental fields live on (degenerate) loops of 
common base point $x$ which are required to be 
invisible at resolution $\La_b$. Because of its chiral definition the connection 
on the loop has vanishing curvature at this resolution. The definition 
of the Higgs field is in analogy to that of the chiral field. 
Section 3 establishes the connection between Higgs VEV and 
the gauge invariant field strength correlator \cite{Dosch}. From this and the use of 
lattice results the scale $\La_b$ is 
evaluated numerically. The largeness of this scale as 
compared to the perturbatively determined 
mass scale $\La_{QCD}$ of the theory may influence the theoretical side of 
QCD sum rules \cite{SVZ} (vacuum average of operator product expansion (OPE)). 
In Section 4 it is argued for 
pure SU(2) gluodynamics (for simplicity) 
that the random condensation of the Higgs field implies the formation of (topologically unstable) 
domain boundaries of positive energy density. This, in turn, 
means that the energy density within the domains must be negative to 
maintain vanishing action. Regions, where three or more domains come together are 
of exceptionally high positive energy and 
therefore exceptionally high instability. From topological arguments we obtain magnetic (anti)monopoles 
at points where four or more domains meet. An immediate consequence of 
the presence of magnetic (anti)monopoles is the breaking of chiral symmetry 
and the proliferation of the axial U(1) 
anomaly at the (large) resolution $\La_b$ when including dynamical 
quarks in the theory. The condensation of 
(anti)monopoles, as it is observed at lower resolution, 
follows from the fact that the Higgs condensate and the (anti)monopole gauge 
fields are weakly varying and the failure to 
resolve the domain walls. (Anti)monopole condensation 
is equivalent to the restoration of apparently broken 
gauge and translational symmetries and goes together with the 
formal masslessness of these objects. Section 5 summarizes the 
results and gives the conclusions.

\section{Definition of vacuum fields}

For the definition of low energy fields, which potentially describe the vacuum, 
we appeal in this section 
to Wilson's ideas about the grain coarsing of spacetime \cite{Wilson} and 
to Polyakov's approach to gauge field dynamics in terms 
of chiral fields \cite{Pol2}. 
At resolution $\La_b$ the partition function $Z_{\La_b}$ 
of a theory, defined in terms 
of local fields $\phi_{\La_b}$ and an action $S_{\La_b}$, is the same as 
the partition function $Z$ at 
resolution $\La\to\infty$, defined in terms of local 
fields $\phi$ and the continuum action $S$
\eqb
\label{RG}
Z_{\La_b}=\sum_{\phi_{\La_b}}\exp\left(-S_{\La_b}\right)=
\sum_{\phi}\exp\left(-S\right)=Z\ .
\eqe
Fields $\phi_{\La_b}$ summed over in $Z_{\La_b}$ 
fluctuate at length scales 
$|x-y|\ge\La_b^{-1}$ and higher. In the case of gauge theories the classical continuum 
action has to be extended 
by a gauge fixing term and the corresponding 
Faddeev-Popov determinant. 
Due to (\ref{RG}) the definition of relevant 
fields at resolution $\La_b$ in terms of continuum fields 
involves some kind of averaging over (euclidean) 
volumes of size $\sim\La_b^{-4}$. Therefore, 
we expect this definition to be nonlocal. Since an analytical grain 
coarsing leading to $S_{\La_b}$ is unfeasible some guess on the definition 
of the relevant fields must be made. Here, we use the 
ideas of ref.\,\cite{Pol2} about stringy objects in gauge 
theories to define local 
fields at resolution $\La_b$. The basic quantity is the (euclidean) 
Wilson loop $W(C)$ of contour $C$ with base point $x$
\eqb
\label{W}
W(C)\equiv{\cal P}\exp[\int_C dy_\mu A_\mu]\ ,
\eqe
where ${\cal P}$ demands path ordering, and the contour $C$ 
is parametrized by a 
dimensionless variable $s$ which may conveniently be chosen to 
measure the normalized length of the curve. Hence, we have  
\eqb
y_\mu(s=0)=y_\mu(s=1)=x_\mu\ .
\eqe
Note that our fundamental gauge field $A_\mu$ is defined to be 
an anti-selfadjoint operator. It is obtained by multiplying the 
gauge field in perturbative definition by $-ig$, where $g$ is the gauge
coupling. 

\noindent The anti-selfadjoint connection $a_\mu(s,C)$ on the loop $C$ is defined to be a chiral field:
\eab 
\label{A}
a_\mu(s,C)&\equiv&\frac{\delta W(C)}{\delta y_\mu(s)}\,W^{-1}(C)\nonumber\\ 
&=&S(x,y(s)) 
F_{\mu\nu}(y(s)) \frac{dy_{\nu}(s)}{ds}\,S(y(s),x)
\eae
where $W^{-1}(C)$ is the Wilson 
loop defined by (\ref{W}) when running 
through the countour $C$ backwards, $F_{\mu\nu}\equiv\pd_\nu A_\mu-\pd_\mu A_\nu+[A_\mu,A_\nu]$, and 
\eqb
S(x,y(s))\equiv{\cal P}\exp[\int_x^{y(s)} dy_\mu A_\mu]\ .
\eqe
Using the Yang-Mills equations, it was shown 
in \cite{Pol2} that $a_\mu(s,C)$ has 
vanishing curvature on the loop
\eqb
\label{cur}
\frac{\delta a_\mu(s,C)}{\delta y_\nu(s^*)}-
\frac{\delta a_\nu(s^*,C)}{\delta y_\mu(s)}+[a_\mu(s,C),a_\nu(s^*,C)]=0
\eqe
for $s^*\le s$. Furthermore, one has
\eqb
\label{div}
\frac{\delta a_\mu(s,C)}{\delta y_\mu(s)}=0\ .
\eqe
To implement grain coarsing we define on the operator level at resolution 
$\La_b$ a local, anti-selfadjoint field $\al_\mu(x)$ as follows:
\eab
\label{al}
\al_\mu(x)&\equiv&\int_{\tiny{\begin{array}{c}{y(0)=x}\\ 
\mbox{max}|y(s)-x|\le\La_b^{-1}\end{array}}}^{\tiny{\begin{array}{c}y(1)=x
\end{array}}}{\cal D}y\,\int_0^1 ds\, a_\mu(s,C)\nonumber\\ 
&=&\int_{\tiny{\begin{array}{c}{y(0)=x}\\ 
\mbox{max}|y(s)-x|\le\La_b^{-1}\end{array}}}^{\tiny{\begin{array}{c}y(1)=x
\end{array}}}{\cal D}y\,\int_0^1 ds\,  S(x,y(s)) 
F_{\mu\nu}(y(s)) \frac{dy_{\nu}(s)}{ds}\,S(y(s),x)\ ,
\eae
where a normalization factor, which gives $\al_\mu$ 
the canonical mass dimension 1, has been absorbed into the 
integration measure ${\cal D}y(s)$. 
So (\ref{al}) defines a local field $\al_\mu(x)$ by integrating the chiral loop 
field $a_\mu(s,C)$ over all loops which 
do not leave a sphere of radius $\La^{-1}_b$ about 
the base point $x$. This is the implementation of the demand that the field 
$\alpha_\mu$ is {\sl local} at resolution $\La_b$. Note that $\alpha_\mu$ transforms 
homogeneously under gauge transfromations of
the fundamental field $A_\mu$. 

In order to make contact with the 
lattice measurement of the gauge invariant field strength correlator \cite{Dosch} 
it will turn out later that we have to constrain the functional 
integration over all loops to an integration 
over (nearly) degenerate loops constituted by (nearly) straight lines connecting the base point $x$ 
with points on the sphere and vice versa. Evaluating the partition function at resolution $\La_b$, 
this seems to be reasonable as long as the relevant configurations\footnote{We do not notationally distinguish between 
operators and classical fields at resolution $\La_b$.} $\alpha_\mu$ do not 
vary much over distances $\sim \La_b^{-1}$. We will denote this constrained 
functional integration by a measure ${\cal D}^\prime y$. 
\begin{figure}
\vspace{6.3cm}
\includegraphics{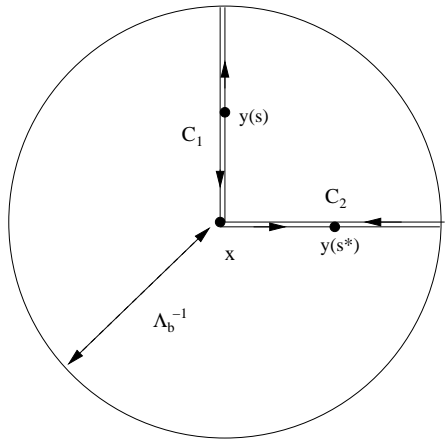}
\caption{A parallel transport arising from the commutator term in (\ref{prop}).} 
\label{} 
\end{figure}
A properties similar to (\ref{cur}) belonging to the loop field $a_\mu(s,C)$ is 
inherited by the local field $\al_\mu(x)$ if only contributions from one 
and the same loop are considered in the commutator term. The analogue of (\ref{div}) is always valid. 
This can be seen as follows: 
On the one hand, we have 
\eab
\label{der}
\tilde{\pd}_\nu \al_\mu(x)&=&\int_{y(0)=x}^{y(1)=x}
{\cal D}^\prime y\,\int_0^1 ds\,\int_0^1 ds^*\, 
\frac{\delta a_\mu(s,C)}{\delta y_\kappa(s^*)}\, \tilde{\pd}_{x_\nu} y_{\kappa}(s^*)\nonumber\\ 
&=&\int_{y(0)=x}^{y(1)=x}
{\cal D}^\prime y\,\int_0^1 ds\,\int_0^1 ds^*\, 
\frac{\delta a_\mu(s,C)}{\delta y_\kappa(s^*)}\delta_{\kappa\nu}\nonumber\\ 
&=&\int_{y(0)=x}^{y(1)=x}
{\cal D}^\prime y\,\int_0^1 ds\,\int_0^1 ds^*\, 
\frac{\delta a_\mu(s,C)}{\delta y_\nu(s^*)}\ .
\eae
On the other hand, disregarding cross 
terms from distinct lines (see Fig. 1) in the commutator $[\al_\mu(x),\al_\nu(x)]$, we have 
\eqb
[\al_\mu(x),\al_\nu(x)]=\int_{y(0)=x}^{y(1)=x}
{\cal D}^\prime y\,\int_0^1 ds\,\int_0^1 ds^*\,[a_\mu(s,C),a_\nu(s^*,C)]\ .
\eqe
Hence, we obtain
\eqb
\label{prop}
\tilde{\pd}_\nu \al_\mu(x)-\tilde{\pd}_\mu \al_\nu(x)+[\al_\mu(x),\al_\nu(x)]=0\ 
,\ \ \ \tilde{\pd}_\mu \al_\mu(x)=0\ ,
\eqe
which shows that $\al_\mu$ is a vacuum field. 
Note the difference between the derivatives $\pd$ and $\tilde{\pd}$. 
The former operates on vanishing distances 
while the latter resolves only distances of the order $\La^{-1}_b$. 

Having 
constructed at resolution $\La_b$ a field $\al_\mu(x)$, 
which is pure gauge, it is natural to ask 
what other field operators relevant for the vacuum description one can define in the above spirit. 
If there is a 
resolution $\La_b$ from which on the vacuum 
drastically changes its appearance, this 
transition must be described 
by a scalar quantity $\phi$. Moreover, imposing Abelian gauges, 
the potential reformulation of the YM action into a 
Georgi-Glashow model, as advertised in ref.\,\cite{Sur}, and an old 
consideration by Kleinert 
\cite{Kleinert} suggest that this scalar field is an adjoint one. Mass dimension four 
is the lowest possible for the definition 
of a scalar in terms of field strength. 
So we define\footnote{We have $F_{\mu\nu}(y(s))\frac{dy_\mu(s)}{ds}\frac{dy_\nu(s)}{ds}=0=
F_{\mu\mu}(y(s))$. Thus a definition of $\phi$ in terms of a single 
field strength tensor and only tangential vectors is impossible. 
On the other hand, a scalar construction 
involving a bilocal product of field strength does not necessitate 
the use of tangential vectors.} 
\eab
\label{phi}
\phi^2(x)&\equiv&\phi^a(x)\phi^b(x)\frac{t_a}{2}\frac{t_b}{2}\nonumber\\ 
&\equiv&-\int_{y(0)=x}^{y(1)=x}
{\cal D}^\prime y\,\int_0^1 ds\,F_{\mu\nu}(x) S(x,y(s)) 
F_{\mu\nu}(y(s)) \,S(y(s),x)\ ,
\eae
where $t_a$ denote the generators of the gauge group 
in the fundamental representation (tr\,$t_at_b=2\delta_{ab}$), and a 
normalization factor, which gives $\phi$ the canonical mass dimension 1, has been absorbed 
into the integration measure ${\cal D}^\prime y$. Note that although already $\phi(x)$ transforms homogeneously 
under gauge transformations of the fundamental fields, as does the r.h.s. of (\ref{phi}), 
it is necessary to have the square of $\phi$ on the l.h.s. 
to avoid a contradiction when taking the 
color trace and the vacuum average 
on both sides of the equation. Definition (\ref{phi}) can be rewritten as
\eqb
\label{phir}
\phi^2(x)=-\lambda(x) \int_{|z-x|\le\La_b^{-1}}d^4z\, F_{\mu\nu}(x) S(x,z) 
F_{\mu\nu}(z) \,S(z,x)\ .
\eqe
The scalar singlet factor $\lambda(x)$ is of mass dimension two. 
Taking the color trace and the vacuum average of the r.h.s. of (\ref{phir}), 
a genuine $x$ dependence of $\lambda$ signals the 
breaking of translational invariance at resolution $\La_b$ 
due to the so-defined classical field configuration $\phi^a\phi^a$. In Section 4 
this will be argued to happen due to the 
{\sl random} condensation of $\phi$.  Note that 
in the vacuum average of the r.h.s. of (\ref{phir}) only physical fluctuations of 
the fundamental fields which are of ``hardness'' $\ge \La_b$ 
contribute in an essential way to the spacetime integral. Only these fluctuations contribute 
sizably to the correlator at the respective distance $|z-x|$. This is a realization 
of Wilson's ``integrating out higher scales''.

\section{Evaluation of the scale $\La_b$}

To estimate the scale $\La_b$ we consider 
the gauge invariant quantity obtained by taking the trace and the 
vacuum average on both sides of (\ref{phir}). The result can be written as follows 
\eab
\label{Fc}
\la\phi^a\phi^a\ra&=&-2\times\lambda(x)\int_{|z-x|\le \La^{-1}_b} d^4z \,\mbox{tr}\,\la F_{\mu\nu}(x) S(x,z) 
F_{\mu\nu}(z) \,S(z,x)\ra\nonumber\\ 
&\equiv&2\times \lambda(x)\int_{|z-x|\le \La^{-1}_b} 
d^4z F_{\mu\nu,\mu\nu}(z^2)\nonumber\\ 
&\equiv&2\times \lambda(x)\int_{|z-x|\le \La^{-1}_b} 
d^4z \left\{12[D(w^2)+D_1(w^2)]+6w^2\pd_{w^2}D_1(w^2)\right\}\ ,
\eae
where $w\equiv z-x$. The last line is due to ref.\,\cite{Dosch} 
where the invariants $D$ and $D_1$ have been introduced in a parametrization of the gauge invariant field strength
correlator $F_{\mu\nu,\mu\nu}(z^2)$. Moreover, 
the same generators $t^a/2$ have been used on the l.h.s. and r.h.s. of (\ref{Fc}).  

Although there has been a very interesting analytical 
investigation of $F_{\mu\nu,\mu\nu}(z^2)$ in terms of vacuum 
fields saturated by constrained instanton solutions \cite{Doro} we appeal to 
direct lattice information in the present work. Measurements of $D$ and $D_1$ on a 
lattice with gauge group SU(3) and $N_F=4$ dynamical fermions of 
the common physical masses $m_q\sim 18$ MeV and $m_q\sim 36$ MeV 
at a lattice resolution of $a\sim 0.11$ fm have been performed in ref.\,\cite{DiGiacomo}. 
We will refer to these cases as (i) and and (ii), respectively. 
In refs.\,\cite{Me} $D$ and $D_1$ were 
measured for pure SU(3) gluodynamics. 
This case will be referred to as (iii). It should be mentioned at this point 
that dynamical quarks to a certain extend spoil the 
zero curvature property of the field $\alpha_\mu(x)$ in pure gluodynamics since the 
{\sl sourceless} Yang-Mills equations have been used to derive (\ref{cur}).

The following functions were fitted:
\eab
\label{Ds}
D(w^2)&=&A_0\exp[-|w|/\lambda_A]+\frac{a_0}{|w|^4}\exp[-|w|/\lambda_a]\ ,\nonumber\\ 
D_1(w^2)&=&A_1\exp[-|w|/\lambda_A]+\frac{a_1}{|w|^4}\exp[-|w|/\lambda_a]\ .
\eae
In terms of an operator product expansion (OPE) of the correlator $F_{\mu\nu,\mu\nu}$ it 
is apparent that the terms in (\ref{Ds}) containing $|w|^{-4}$ are chiefly due 
to renormalon-free (that is unsummed) 
perturbative contributions\footnote{To zeroth order in $\alpha_s$ one obtains 
$D(w^2)=0$ and $D_1(w^2)=\frac{4g^2}{\pi^2 |w|^4}$ \cite{eidemuller}.}\cite{DiGiacomo}. Since the 
formation of composites is a genuinely 
nonperturbative feature and 
since the definition (\ref{Fc}) would be ill if one 
carried along the power-law like behavior we restrict ourselves 
to the purely exponential decay of the correlator. It is an easy exercise to perform 
the $z$ integration in (\ref{Fc}). The result is
\eab
\label{Fcc}  
\int_{|z-x|\le \La^{-1}_b} d^4z F_{\mu\nu,\mu\nu}(z^2)&=&2\pi^2\left\{12[A_1+A_0]\left(6\lambda_A^4-
\lambda_A\exp[-1/(\La_b\lambda_A)](\La_b^{-3}+
3\La_b^{-2}\lambda_A+6\La_b^{-1}\lambda^2_A +6\lambda^3_A)\right)-\right.\nonumber\\ 
& &\left.3\frac{A_1}{\lambda_A}\left(24\lambda_A^5-\lambda_A\exp[-1/(\La_b\lambda_A)](\La_b^{-4}+
4\La_b^{-3}\lambda_A+12\La_b^{-2}\lambda^2_A +24\lambda^4_A)\right)\right\}\ .\nonumber\\ 
\eae
To estimate the scale $\La_b$ we demand
\eqb
\label{self}
\la \phi^a\phi^a\ra(x)\equiv\rho=2\times\lambda(x)\int_{|z-x|\le \La^{-1}_b} d^4z\, F_{\mu\nu,\mu\nu}\ ,
\eqe
where due to the translationally invariant integral in (\ref{self}) 
$\rho(x)$ is proportional to $\lambda(x)$. The ambiguity of the local 
magnitude of $\lambda(x)$ in the definition (\ref{phi}) is then parametrized in terms of 
the dimensionless quantity $\xi\equiv\frac{\rho}{\lambda}$. By means of 
(\ref{self}) we obtain the following implicit definition of the function 
$\La_b=\La_b(\xi)$
\eqb
\label{imp}
f(\xi,\La_b(\xi))\equiv\xi-2\times\int_{|z-x|\le \La^{-1}_b} d^4z F_{\mu\nu,\mu\nu}=0\ .
\eqe
To numerically evaluate this equation we take the 
central values from \cite{DiGiacomo} and \cite{Me} 
for the parameters $\lambda_A^{-1},A_1,A_0$ which, when expressed 
in physical units, are
\eab
(i):\ \ \lambda_A^{-1}=0.588\, \mbox{GeV}\ ,\ \ &A_0=2.367\times 10^{-2}\,\mbox{GeV}^4&\ ,\ 
\ A_1=2.72\times 10^{-3}\,\mbox{GeV}^4\ ;\nonumber\\ 
(ii):\ \ \lambda_A^{-1}=0.689\, \mbox{GeV}\ ,\ \ &A_0=4.966\times 10^{-2}\,\mbox{GeV}^4&\ ,\ 
\ A_1=6.49\times 10^{-3}\,\mbox{GeV}^4\ ;\nonumber\\ 
(iii):\ \ \lambda_A^{-1}=0.895\, \mbox{GeV}\ ,\ \ &A_0=1.957\times 10^{-1}\,\mbox{GeV}^4&\ ,\ 
\ A_1=4.1\times 10^{-2}\,\mbox{GeV}^4\ . 
\eae
The first observation is that for all cases (i)--(iii) the large range of $\xi$ values $0.2\le\xi\le 5.0$ 
there is a {\sl unique} zero of $f(\xi,\La_b)$ in $\La_b$. Second, the function $\La_b(\xi)$ 
varies rather slowly (see Fig. 2). For example, for (i) a factor 25 in 
$\xi$ implies a factor $\sim$ 3.1 in $\La_b$. Apart from the 
arguments of Section 2, this seems to give additional support to 
the definition (\ref{Fc}). 
\begin{figure}
\vspace{7.9cm}
\includegraphics{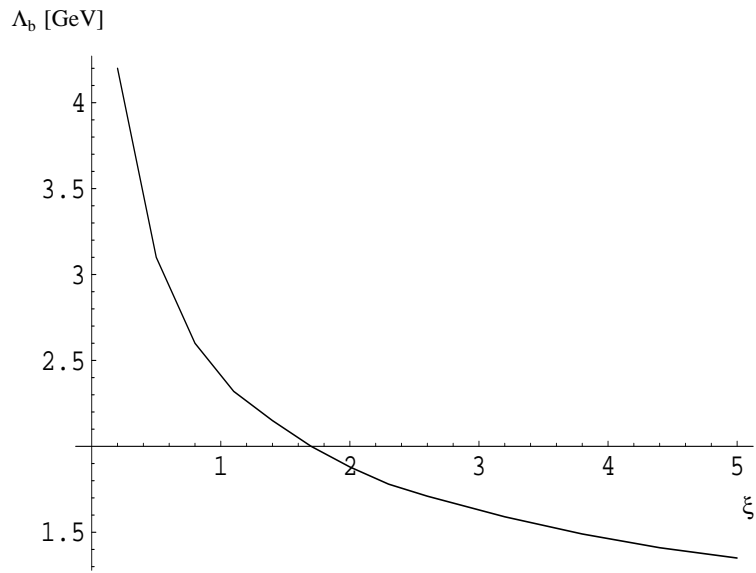}
\caption{The resolution scale $\La_b$ as a function of $\xi$ for the case (i).} 
\label{} 
\end{figure}
Naturally, the formation scale of 
a Higgs condensate roughly coincides with the value of this condensate. We will use 
this fact in Section 4 to argue that Abelian projection 
is dynamical for resolutions $\le\La_b$. Resorting to 
$\xi\sim 1$ \footnote{This corresponds to the demand that 
Higgs condensation is described by a single scale $\La_b$.}, we have 
\eqb
\label{res}
(i):\ \La_b(\xi\sim 1)\sim 2.4\,\mbox{GeV}\,;\ \ \ (ii):\ \La_b(\xi\sim 1)\sim  3.1\,\mbox{GeV}\,;
\ \ \ (iii):\ \La_b(\xi\sim 1)\sim 4.2\, \mbox{GeV}\ .
\eqe
The most realistic case (i) corresponds to a condensation scale which is much 
higher than perturbatively determined scales $\La_{QCD}$ which, depending on the 
number of active quark flavors and the renormalization 
scheme, range between 200--500 MeV. It is tempting to extrapolate to more 
realistic, that is, smaller quark masses. However, since $N_F=4$ and 
{\sl equal} quark masses were assumed we feel that the uncertainty 
of such an extrapolation is too high for a conclusive statement. 
Hopefully, the results of more realistic lattice data 
simulations will soon become available. 
From (i) and (ii) in (\ref{res}) it is clear that 
even at small quark mass $m_q$ the condensation scale $\La_b$ depends 
rather strongly on it.

If the condensation scale $\La_b$ in realistic QCD is higher 
than 1 GeV one would expect the standard 
operator product expansion (OPE) used in QCD sum rules \cite{SVZ} to be affected.
At this scale one usually evaluates QCD sum rules \cite{SVZ}. There, the power corrections to 
a gauge invariant current correlator, introduced via the vacuum average of the corresponding 
OPE, start at dimension four. This is a consequence of the {\sl perturbative} calculation of 
Wilson coefficients. If there is a condensation of a Higgs field 
at the typical sum rule scale, as it is suggested by the above, the 
description of the non-perturbative QCD vacuum solely 
in terms of averages over local operators constructed from fundamental 
fields may not be sufficient since a composite scalar $\phi$ creates power 
corrections $\sim \la \phi^a\phi^a\ra/Q^2$. 
In other words, the separation of nonlocal, potentially non-perturbative short distance 
effects into perturbatively calculated Wilson coefficients, which 
yield {\sl logarithmic} factors\footnote{Note, however, that the summation of 
certain perturbative diagrams leads to ambiguities (UV renormalons) which 
formally cause dimension two power corrections \cite{Beneke}.}, 
is then not guaranteed to lead to an adequate discription 
at resolution $Q\sim 1$ GeV. In refs.\,\cite{Gub1} 
and \cite{Shu} power corrections due the short distance 
dynamics of small-size strings and instantons, respectively, have been argued 
to exist. In ref.\,\cite{Nar-Zakh} the introduction of $Q^{-2}$ power corrections 
due to non-perturbative short-distance effects 
was based on phenomenological grounds using the concept of a tachyonic gluon mass $\lambda$. 
It was found that a value as high as $|\lambda|=0.7$ GeV is compatible with 
phenomenology in a variety of channels ($\rho, \pi$, scalar gluonium). 
Another approach to a short-distance dimension-two
condensate in QCD defined in terms of fundamental fields 
has been proposed in refs.\,\cite{Gub}. On the other hand, 
it was shown in ref.\,\cite{Nar-Zakh} 
that the ad hoc subtraction of the Landau pole 
in the running coupling $\alpha_s(Q^2)$ leads 
to a soft power correction $\sim\Lambda^2_{QCD}/Q^2$. 
Attempts to develop a phenomenology of these long distance, power-two 
corrections were reported in refs.\,\cite{ld}. It is stressed at 
this point that the formation of a Higgs 
condensate at resolution $\La_b$ {\sl freezes} the perturbative 
running of the coupling to the value at this scale. So in the context of the present work 
there is a more profound alteration of the running of 
the coupling than a simple subtraction of the Landau pole.

\section{How magnetic monopoles may form and condense}

Based on the above in this section we try to 
understand how the dual Meissner effect 
may be realized. In order to make the discussion simple 
we first consider pure SU(2) gluodynamics. Generalizations to SU(3) 
are straightforward.   

Taking it as a fact that the vacuum below resolution $\La_b$ is characterized by the 
dynamics of a Higgs field $\phi^a$ and a field $\al_\mu$, which is pure gauge, 
one may construct a Georgi-Glashow like model with the 
curvature term for the gauge field missing:
\eqb
\label{GG}
{\cal L}_{vac}=\frac{1}{2} D_\mu\phi^a D_\mu\phi^a-V(\phi^a\phi^a)\ .
\eqe
Thereby, $D_\mu\phi^a\equiv \tilde{\pd}_\mu \phi^a-ig\ep^{abc}\al_\mu^b\phi^c$, 
and $V$ is some gauge invariant potential. Note that the connection $\alpha_\mu$ in 
this section is defined according to the perturbative convention, 
that is, one obtains $\alpha_\mu$ used here by 
multiplying $\alpha_\mu$ of Section 2 with $i/g$. In the background of $\alpha_\mu$-fluctuations
close to zero the apparent formation of a scalar condensate at resolution $\La_b$ 
proceeds from condensation seeds. With these seeds present the growth 
of (3-dimensional) bubbles of constant Higgs field (constant color direction and modulus) 
sets in. Thereby, the color orientation inside a particular bubble is 
{\sl unrelated} to that in the neighbouring bubble. Eventually, bubble edges collide. 
The probability that the Higgs directions of 
neighbouring bubbles coincide is exactly zero. Consequently, 
there are "discontinuities" of the color orientation 
of the Higgs field over length scales of the order of the resolution $\La_b^{-1}$ 
across the bubble boundary. From the kinetic part of 
(\ref{GG}) we have finite and positive 
energy density $\ep$ along the boundaries. This can most easily be seen as follows: 
Imposing the unitary gauge $\phi^a=|\phi|\delta^{3a}$ 
globally, one can shift the "discontinuity" of the Higgs direction 
into a "discontinuity" of the field $\al_\mu$\footnote{Within the collision zone the homogenity of the fields $\alpha_\mu$ and
$\phi$ is destroyed. The contributions of undegenerated loops should become important, and $\alpha_\mu$ is lifted from zero to a
nonvanishing pure gauge configuration.}. Without restriction of 
generality we assume that the boundary 
between two neighbouring 
bubbles $A$ and $B$ lies in the $x_1,x_2$-plane and that the Higgs field in $A$ 
already is $\phi^a=|\phi|\delta^{3a}$. The gauge transformation $\Omega$
reaching the global gauge $\phi^a=|\phi|\delta^{3a}$ in the vicinity of the boundary is given as
\eqb
\label{Om}
\Omega=\theta_{\La_b}(-x_3){\bf 1}+\theta_{\La_b}(x_3) U_B\ ,\ \ U_B\equiv i n_\kappa \tau^{\pm}_\kappa\ ,
\eqe
where $\tau^{\pm}=({\vec t}/2,\mp i{\bf 1})$, the components of $n$ ($n_\kappa n_\kappa=1,\,\kappa=1,2,3,4$) 
can be determined from the 
direction of the Higgs field in $B$, and $\theta_{\La_b}$ denotes 
a softened theta function for resolution $\La_b$
\eqb
\label{th}
\theta_{\La_b}(x_3)\equiv\theta(x_3+\La_b^{-1}/2)\,\theta(-x_3+\La_b^{-1}/2)\,\La_b x_3+\theta(x_3-\La_b^{-1}/2)\ .
\eqe
The corresponding $\delta_{\La_b}$ function can be obtained from $\theta_{\La_b}$ 
by differentiation and omission of terms 
which contain ordinary delta functions as factors. 
Using (\ref{Om}), (\ref{th}) and the 
fact that $\al_\mu$ is pure gauge, 
one obtains 
\eab
\label{ep}
\ep&=&\frac{1}{2} \,g^2|\phi|^2\left\{\left(\al^1_\mu\right)^2+\left(\al^2_\mu\right)^2\right\}\nonumber\\ 
&=&\frac{1}{2} \,|\phi|^2(\delta_{\La_b})^2\left(n_1^2+n_2^2\right) 
=2\La_b^4\left(n_1^2+n_2^2\right)\ ,\ \ (-\La_b^{-1}/2\le x_3\le\La_b^{-1}/2)\ ,
\eae
where in the last line 
we have set the Higgs field modulus 
equal to the scale $\La_b$. Note that $n_3$ (or $n_4$) is cyclic 
in (\ref{ep}) which signals the residual U(1) symmetry. 

The vacuum manifold of the theory is the coset space $M=SU(2)/U(1)$. 
It has the topology of a 2-sphere, and therefore it is simply connected. 
Two points can always be continuously connected, and hence 
there is no topological stabilization of the boundary between two domains. 
This means that localized energy 
density along the boundary at the time of collision can 
flow apart \cite{Kibble}. However, this happens due to approaching Higgs directions 
in a vicinity of the collision zone. At another boundary of the bubble, say, $A$ the same process occurs 
but there with the Higgs 
field pointing in a different direction. After some time this implies localization of 
positive energy along a new domain boundary. 

Now the positive energy along the domain boundaries must be countered by negative 
energy inside the domains to yield a vanishing action of the configuration. Negative energy 
is plausible since the formation of a Higgs 
condensate goes with a reduction of entropy density. This 
difference in energy density between the vacuum 
resolved at long and short distances was already incorporated 
in the bag models of hadrons \cite{Chodos74a} by means of the bag 
constant $B$. Recall that the introduction of $B$ is necessary 
to cure the non-conservation of bag four-momentum which is due to the violation of 
translational invariance by the bag boundary. We will 
soon argue that the presence of 
negative energy inside the 
domains, represented by a weakly 
varying Higgs condensate, also enforces a restoration of the (classical)   
translational and gauge symmetry at lower resolution.       
\begin{figure}
\vspace{7.1cm}
\includegraphics{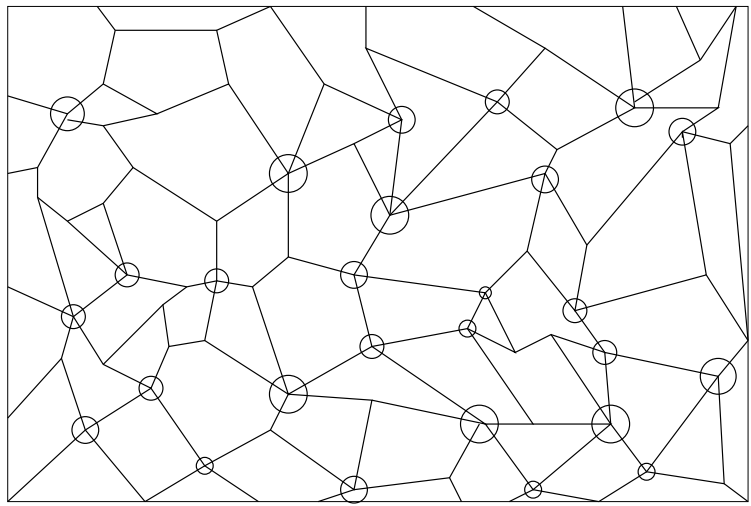}
\caption{A potential snapshot for a spatial slice of the YM vacuum at resolution $\La_b$. The 
solid lines indicate domain boundaries. Regions of potential monopole formation 
are encircled.} 
\label{} 
\end{figure}
So far we only considered the boundary between {\sl two} domains. Regions of 
exceptionally high, positive energy and therefore exceptionally high 
instability of the Higgs condensate are those where several domains come together. 
They can be string-like (3 domains meet along one edge) or 
point-like (generically 4 domains meet). As in the case of two colliding domains 
strings are not stable topologically since any closed curve 
on $M$ can be continuously deformed to a point \cite{Kibble}. 
In the case of four or more domains coming together 
one cannot shrink the corresponding surface in $M$ to a point 
(the second homotopy group $\pi_2(M)$ is non-trivial). 
So the cores of these regions are topologically 
stable \cite{Kibble} and are characterized by a vanishing Higgs condensate. 
The more walls come together in a small 
region the more profound the destruction 
of the Higgs condensate in that region is. The entire gauge group is restored there, 
and we have magnetic monopoles 
or antimonopoles\cite{tH3}\footnote{The properties (\ref{prop}) of 
the gauge field $\al_\mu$ far away from the cores 
are satisfied asymptotically also by the classical gauge field of the 
't Hooft/Polyakov (anti)monopole \cite{tH3}.}. In Fig. 2 a snapshot of a spatial slice of 
the domain structure of the vacuum at resolution $\La_b$ 
is depicted. 

An immediate consequence of the formation of magnetic monopoles and 
antimonopoles has been pointed out in the work of ref.\,\cite{Stern}. 
There, it was argued that static, zero magnetic 
charge configurations such as the 
superposition of a 't Hooft/Polyakov 
monopole and an antimonopole induce a nonvanishing spectral 
density of normalizable zero modes in the 
representation of the massless fermionic 
propagator in this background. According to \cite{Stern} this implies 
a {\sl flavor independent} breaking of chiral symmetry. This is in contrast to 
chiral symmetry breaking due to the instanton. Moreover, the nonvanishing 
magnetic charge {\sl density} is essential for the 
proliferation of the axial U(1) anomaly which prohibits 
the corresponding massless meson in the spectrum \cite{Stern}. So already at 
resolution $\La_b$, which is much higher than the confinement scale, chiral symmetry breaking 
and the effects of the axial U(1) anomaly set in. 

What about the condensation of monopoles? From the above it is clear  
that the average core sizes $R$ of the (anti)monopoles 
are larger than $\La^{-1}_b$ which is comparable to the width of 
the domain boundaries. Furthermore, 
the long-range gauge field of the (anti)monopoles and the 
weak variation of the Higgs 
condensate inside the domains imply that these fields survive a 
grain coarsing to lower resolution. 
This is {\sl not} the case for the domain boundaries. As the 
resolution is tuned down the walls apparently cease to exist. To keep the 
configuration at vanishing action in this process the domains must shrink. 
At a critical resolution 
$\La_c^{-1}<R$ the walls vanish entirely, 
but not yet the cores of the monopoles. The only way to 
keep the action approximately at zero is a situation 
where the cores are arbitrarily close to one 
another to produce a vanishing energy density in the vacuum. This must be 
interpreted as the condensation of magnetic
charges. Since monopoles are reduced to their 
cores they then can be considered massless. 
The relevant configurations at resolution $\le\La_c$ are translationally invariant, and the 
gauge symmetry is restored everywhere since the 
Higgs field vanishes globally. Phenomenologically, one expects 
$\La_c\sim\La_{QCD}\sim 0.3$ GeV. So, 
in agreement with a statement by Polyakov \cite{Pol}, 
the phenomenon of quark confinement goes together with the 
restoration of an apparently broken gauge symmetry. 

To conclude this section let us remark 
on the relation between Abelian projection and 
the above picture in QCD. So far we have 
only considered the vacuum fields of the (anti)monopoles 
far away from a core or inside a core. 
At intermediate distance from the core 
the constraint of nonvanishing curvature for the gauge field might be 
slightly violated, and gauge field 
fluctuations may propagate. Moreover, the presence of 
dynamical quarks also causes the curvature to be 
nonvanishing as was pointed out in the last section. In the following we 
speculate that $\La_b\sim$ 1 GeV in QCD. It will be clear that 
qualitative implications do not depend on the exact value of $\La_b$.   
 
In unitary gauge, $\phi^a=\frac{1}{\sqrt{2}}(
\delta^{3a}+\delta^{8a})|\phi|$, off-diagonal gluons aquire a mass
\eqb
\label{mass} 
m_{W}=g(\La_b)|\phi|\sim \sqrt{4\pi\alpha_s(\La_b)}\La_b\ .
\eqe
Thereby, the running of $\alpha_s$ is frozen at the value $\alpha_s(\La_b)$ for 
resolution lower than $\La_b$. 
Diagonal gluons (photons) remain massless. Taking $\alpha_s(\La_b=1 \mbox{GeV})=0.4$
\footnote{This value stems from a 
determination in the $\overline{\mbox{MS}}$ scheme with 
four-loop evolution and three-loop matching at 
the flavor thresholds
\cite{Potter}.}, one obtains $m_W=2.24$ GeV. So at resolution $\le$1 GeV 
the dynamics of propagating gluons 
is Abelian to a very high degree. Moreover, photons 
are Debye-screened on distances comparable to 
the average monopole separation. At 
resolutions $\sim \La_c$ this distance vanishes and the (anti)monopoles 
become massless, so photons do not propagate at all. To estimate the 
mass of the two neutral Higgs particles let us assume 
that the dynamics of domain creation 
is goverened by a standard Higgs potential
\eqb
V(\phi^a\phi^a)=\beta(\phi^a\phi^a-\La_b^2)^2-\gamma\ .
\eqe
To fix the parameters $\beta,\gamma$ we impose
\eqb
V(0)=0\ ,\ \ \ V(\La_b,\La_b)=-B\ ,
\eqe
where $B$ is the bag constant. This gives $\beta=B/(2\La_b^4)$ and $\gamma=B$. Moreover, the mass 
of the neutral scalars is $m_H=2\sqrt{B}/\La_b$. 
In ref.\,\cite{Hofmann} the bag 
constant $B\sim 0.02$ GeV$^4$ was determined 
from first principles of the bag 
model and by using the QCD trace 
anomaly. We obtain $m_H\sim 280$ MeV. According to this rough estimate, 
from a kinematical point of view these neutral scalars can 
be excited and propagate. However, they must 
be radiated off massive, highly virtual gluons 
which makes their appearance very unlikely. Nevertheless, it is 
tempting to speculate on the potential contribution of neutral Higgs excitations 
to the dark matter content of the universe. This issue 
will not be investigated further here.

\section{Summary and Conclusions}

Let us summarize the results of this work: Inspired by 
general renormalization group arguments \cite{Wilson} and the recently 
advocated expressibility of Abelian gauge fixed gluodynamics in terms of a
Georgi-Glashow-like model \cite{Sur} we have used Polyakov's 
ideas \cite{Pol2} about chiral  fields on the
loop to define the corresponding vacuum fields in terms of fundamental fields 
in a nonlocal way. Appealing to lattice information on the gauge 
invariant field strength correlator \cite{Dosch,DiGiacomo,Me}, the resolution $\La_b$ at which 
Higgs condensation can be observed in the 
vacuum was estimated for theories 
with and without dynamical fermions. 
It turned out that the values of $\La_b$ are much larger 
than the perturbatively determined scales $\La_{QCD}$ of the respective theories. 
This may have implications on the evaluation of OPE vacuum averages in the 
framework of QCD sum rules. Conventionally, Wilson coefficients, which 
carry nonlocal information 
on the scale of the external momentum probing the vacuum (usually $\sim$ 1 GeV), are calculated perturbatively. 
Our results seem to indicate, however, that the 
vacuum behaves highly 
non-perturbatively at this resolution already. Hence, the validity of a perturbative 
expansion of Wilson coefficients is questionable. This was stated in a similar way already 
in ref.\,\cite{Dosch}. 

Subsequently, it was argued that the randomness of the Higgs condensation implies 
the formation of domains which are bounded by positive energy walls. These walls are not stable and subject 
to constant rearrangement. For the action to 
remain zero it is necessary to have negative energy inside the domains. 
At locations where four or more domains meet the Higgs 
condensate is destroyed, and one has (topologically) stable magnetic (anti)monopoles. 
Together with the results of ref.\,\cite{Stern} 
this implies that the onset of chiral symmetry breaking and the 
proliferation of the axial U(1) 
anomaly happens at the large resolution $\La_b$ already. 
Describing the vacuum at resolution $\La\le\La_b$ 
in terms of a chiral gauge field and a 
Higgs field, the condensation of (anti)monopoles at $\La_c<\La_b$ 
is a consequence of the weakly varying Higgs condensate, the long-range vector field 
of the (anti)monopole, and the fact that the width of the 
domain walls is smaller than the size of the cores of (anti)monopoles. 
Potentially propagating gluonic excitations were shown 
to be Abelian in nature. At resolution $\La_c$ these 
excitation are completely Debye-screened 
due to the masslessness and the 
vanishing distance between (anti)monopoles. At higher resolutions 
there is a finite screening length comparable 
to the average distance between (anti)monopoles. On the other hand, 
the propagation of neutral Higgs excitations 
over large distances can not be excluded. 
However, the excitation of such 
particles seems to be extremely unlikely. Further quantitative investigations 
are needed to decide whether these Higgs particles can 
contribute to the dark matter content of the universe in a
sizable way.

\section*{Note added}

After completion of this work the author has become aware of a 
potential identification of the field $\phi/\La_b$ with the field $\hat{n}$ used in the 
nonlinear sigma model approach of Refs.\cite{Fad}. 
The author would like to thank L. D. Faddeev 
for the corresponding remark.

\section*{Acknowledgments}

The author would like to thank the following people for useful conversations and 
discussions: O. Grandjean, F. V. Gubarev, A. Hoang, B. P{\"o}tter, 
M. I. Polikarpov, G. Schierholz, L. Stodolsky, P. van Baal, P. Weisz, and V. I. Zakharov. The author 
is indebted to V. I. Zakharov for a critical reading of the 
manuscript and the hinting on useful literature.

\bibliographystyle{prsty}

\begin{thebibliography}{10}

\bibitem{QCD}
H. Fritzsch, M. Gell-Mann, and H. Leutwyler, Phys. Lett. B {\bf 47},  365
  (1973).\\ 
D.~J. Gross and F. Wilczek, Phys. Rev. {\bf D} {\bf 8},  3633  (1973).\\ 
H.~D. Politzer, Phys. Rev. Lett. {\bf 30},  1346  (1973).

\bibitem{HM}
S. Mandelstam, Phys. Rep. C{\bf 23}, 245 (1976).\\ 
G. 't Hooft, Nucl. Phys. B{\bf 190}, 455 (1981). 

\bibitem{'thooft2}
G. 't Hooft, Nucl. Phys. B{\bf 190}, 455 (1981)

\bibitem{Brower}
R. C. Brower, K. N. Orginos, and C.-I. Tan, Phys. Rev. D{\bf 55}, 6313 (1997).\\ 
F. Bruckmann, T. Heinzl, T. Vekua, A. Wipf, Nucl. Phys. B{\bf 593}, 545 (2001).\\ 
R. Hofmann, hep-ph/0101317. 

\bibitem{Kleinert}
H. Kleinert, Lett. Nuov. Cim. {\bf 34}, 209 (1982).

\bibitem{Sur}
P. Suranyi, hep-lat/0102009

\bibitem{Dosch}
H. G. Dosch, Phys. Lett. B {\bf 190}, 177 (1987).\\ 
H. G. Dosch and Yu. A. Simonov, Phys. Lett. B{\bf 205}, 339 (1988).

\bibitem{Wilson}
K. G. Wilson and J. Kogut, Phys. Rept.{\bf 12}, 75 (1974).

\bibitem{Pol2}
A. M. Polyakov, Nucl. Phys. B{\bf 164}, 189 (1980)

\bibitem{Doro}
A. E. Dorokhov, S. V. Esaibegian, A. E. Maximov, and S. V. Mikhailov, 
Eur. Phys. J. C{\bf 13}, 331 (2000), hep-ph/9903450
    
\bibitem{DiGiacomo}
M. D'Elia, A. Di Giacomo, and E. Meggiolaro, Phys. Lett. B{\bf 408}, 315 (1997).

\bibitem{Me}
A. Di Giacomo and H. Panagopoulos, Phys. Lett. B{\bf 285}, 133 (1992).\\ 
A. Di Giacomo, E. Meggiolaro, and H. Panagopoulos, Nucl. Phys. B{\bf 483}, 371 (1997).

\bibitem{eidemuller}
M. Eidem\"uller and M. Jamin, Phys. Lett. B{\bf 416}, 415 (1998).     

\bibitem{Beneke}
M. Beneke, Phys.Rept.{\bf 317},1 (1999).

\bibitem{SVZ}
M. Shifman, A. Vainshtein, and V. Zakharov, Nucl. Phys. B{\bf 147}, 385 (1979).

\bibitem{Nar-Zakh}
K. G. Chetyrkin, S. Narison, and V. I. Zakharov, Nucl. Phys. B{\bf 550}, 353 (1999).

\bibitem{Gub}
F.V. Gubarev, L. Stodolsky, and  V.I. Zakharov, hep-ph/0010096.\\ 
F.V. Gubarev and V.I. Zakharov, hep-ph/0010096. 

\bibitem{Gub1}
F. V. Gubarev, M. I. Polikarpov, and V. I. Zakharov, Mod. Phys. Lett. A {\bf 14}, 2039 (1999)

\bibitem{Shu}
E. V. Shuryak, hep-ph/9909458

\bibitem{ld}
V. I. Zakharov, Nucl. Phys. B{\bf 385}, 452 (1992).\\ 
S. Narison, Phys. Lett. B {\bf 300}, 293 (1993).\\ 
C. A. Dominguez, Phys. Lett. B {\bf 345}, 291 (1995).\\ 
A. I. Vainshtein and V. I. Zakharov, Phys. Rev. Lett. {\bf 73}, 1207 (1994); 
Phys. Rev. {\bf D} {\bf 54},  4039  (1996).\\ 
G. Altarelli, P. Nason, and G. Ridolfi, Z. Phys. C{\bf68}, 257 (1995).\\ 
K. Yamawaki and V. I. Zakharov, hep-ph/9406399, hep-ph/9406373.\\ 
S. Peris and E. de Rafael, Nucl. Phys. B{\bf 500}, 325 (1997).\\ 
M. Beneke, V. M. Braun, and N. Kivel, Phys. Lett. B {\bf 404}, 315 (1997).     

\bibitem{Kibble}
T. W. B. Kibble, J. Phys. A{\bf 9}, 1387 (1976).


\bibitem{Chodos74a}
A. Chodos {\it et~al.}, Phys. Rev. {\bf D} {\bf 9},  3471  (1974).\\ 
T. DeGrand, R.~L. Jaffe, K. Johnson, and J. Kiskis, Phys. Rev. {\bf D} {\bf 
12},  2060  (1975).\\ 
A. Thomas, Adv. Nucl. Phys. {\bf 13},  1  (1984).\\ 
R. Hofmann, M. Schumann, and R.~D. Viollier, Eur. Phys. J. C {\bf 11},  153 (1999).\\ 
R. Hofmann, M. Schumann, T. Gutsche, and R. D. Viollier, Eur. Phys. J. C {\bf 16}, 677 (2000). 


\bibitem{tH3}
G. 't Hooft, Nucl. Phys. B{\bf 79}, 276 (1974).

\bibitem{Stern}
E. Florates and J. Stern, Phys. Lett. B {\bf 119},  419 (1982).

\bibitem{Pol}
A. M. Polyakov, Nucl. Phys. B {\bf 120},  429  (1977).

\bibitem{Potter}
B A. Kniehl, G. Kramer, and B. P{\"o}tter, Phys. Rev. Lett. {\bf 85}, 5288 (2000).\\ 
T. van Ritbergen, J. A. M. Vermaseren, and S. A. Larin, Phys. Lett. B {\bf 400}, 379 (1997).\\ 
K. G. Chetyrkin, B. A. Kniehl, and M. Steinhauser, Phys. Rev. Lett. {\bf 79}, 2184 (1997). 


\bibitem{Hofmann}
R. Hofmann, M. Schumann, T. Gutsche, 
and R. D. Viollier, Eur. Phys. J. C {\bf 16}, 677 (2000). Phys. Rev. Lett. {\bf 82}, 1624 (1999).\\ 


\bibitem{Fad}
L. D. Faddeev and A. J. Niemi, Phys. Rev. Lett. {\bf 82}, 1624 (1999).\\  
L. D. Faddeev, Princeton Report No. IAS-75-QS70, 1970.\\ 
L. D. Faddeev and A. J. Niemi, hep-th/0101078.  


\end{thebibliography}

\end{document}